# REESSE1+ · Reward · Proof by Experiment · A New Approach to Proof of P ≠ NP [*]

Shenghui Su [1] and Shuwang Lü [2]

[1] College of Computers, Beijing University of Technology, Beijing 100124, P. R. China
[2] Graduate School, Chinese Academy of Sciences, Beijing 100039, P. R. China

**Abstract**: The authors discuss what is provable security in cryptography. Think that provable security is asymptotic, relative, and dynamic, and only a supplement to but not a replacement of exact security analysis. Because the conjecture P ≠ NP has not been proven yet, and it is possible in terms of the two incompleteness theorems of Kurt Gödel that there is some cryptosystem of which the security cannot or only ideally be proven in the random oracle model, the security of a cryptosystem is between provability and unprovability, and any academic conclusion must be checked and verified with practices or experiments as much as possible. Extra, a new approach to proof of P ≠ NP is pointed out. Lastly, a reward is offered for the subexponential time solutions to the three REESSE1+ problems: MPP, ASPP, and TLP with $n \geq 80$ and $\lg M \geq 80$, which may be regarded as a type of security proof by experiment.

## 1 Proof and Provable Security

Shimon Even said, "A proof is whatever convinces me." [1]

Oded Goldreich, a cryptologist, thinks that traditionally in mathematics, a "proof" is a fixed sequence consisting of statements which either are self-evident or are derived from previous statements via self-evident rules, and however, in other areas of human activity, the notion of a "proof" has a much wider interpretation [1].

The several types of proof methods in common use are proof by construction, proof by contradiction, proof by induction, proof by deduction or reduction, and proof by combination of the preceding methods [2].

*Security is a type of attribute of a cryptosystem or protocol.*

A cyptosystem or protocol is said to have provable security if its partial or all security requirements can be stated formally in an antagonistic model, as opposed to heuristically, with clear assumptions that certain computational problems are arduous, and the adversary has access to the cryptosystem as well as enough computational resources [3].

The proof of security (called a "reduction") is that these security requirements are met provided the assumptions about the adversary's access to the cryptosystem are satisfied and some clearly stated assumptions about the hardness of certain computational problems hold [3].

There are several approaches to provable security. One is to establish the "correct" definition of security for a given and intuitively understood computational task or problem. Another is to suggest constructions and reasonings based on general assumptions as much as possible. For example, the holding of the conjecture P ≠ NP. Some are in given theoretical models such as the random oracle model, where real hash functions are replaced by an idealization [3].

No real function can implement a true random oracle. In fact, it is known that certain artificial signature and encryption schemes are proven secure in the random oracle model, but are trivially insecure when any real function is substituted for the random oracle [4].

It shouldn't be given indifference that recently Koblitz and Menezes have criticized aspects of provable security in their papers *Another Look at "Provable Security"* and *Another Look at "Provable Security" II*. These views have been controversial in the community. Very recently AMS published a controversial article by Koblitz titled *"The Uneasy Relationship Between Mathematics and Cryptography"*. Several rebuttals have been written and posted [6][7].

---

[*] Manuscript first received Aug. 4, 2009, and last revised Aug. 25, 2014.
Corresponding e-mails: reesse@126.com.
The offer is supported by JUNA.

## 2 Provable Security Is Asymptotic, Relative, and Dynamic — A Cryptosystem Is between Provability and Unprovability

The provable security of a cryptosystem or a digital signer is asymptotic, relative, and dynamic, which can be understood because

① the proof of security of a cryptosystem or a digital signer is through reduction, which decides that the security is relevant to some intractabilities;

② the one-wayness of an intractability is based on computational complexity, which indicates that the security is asymptotic from the threshold value of a dominant parameter;

③ the security proof is made always with some assumptions — an interger factorization problem having no polynomial time solution for example, yet they are asymptotic, relative, and dynamic;

④ the speed of a computer has been rising continually since 1978;

⑤ the model of computation will lift to the quantum Turing machine or others;

⑥ if provable security can assure that a scheme is absolutely or unconditionally secure, a pair of "spears" and "shields" will disappear, and cryptanalysts will lose their occupations.

In visualization, iron doors are firmer than wood doors from an asymptotic or relative sense, but we may not say that an iron door 0.5 mm thick is firmer than a wood door 10 mm thick. An iron door 10 mm thick is secure now, but we may not say that the iron door 10 mm thick is still secure in future.

In provable security theory, the most basic assumption $P \neq NP$ has not been proven yet. Additionally, in terms of the two incompleteness theorems of Kurt Friedrich Gödel [8], it is possible that the security of some cryptosystem, especially some multivariate or multiproblem cryptosystem, cannot or only ideally be proven in the random oracle model. Gödel essentially constructed a formula which claims that it is unprovable in a given formal system because if it were provable, it would be false, which contradicts the idea that in a consistent system, provable statements are always true [8]. Thus, the security of a cryptosystem is right between provability and unprovability.

## 3 Provable Security Is Only a Supplement to But Not a Replacement of Security Analysis

The significance of provable security consists in the thing that it provides a piece of theoretical evidence that a cryptographic scheme should be secure or a compuational problem should be intractable generically; nevertheless, it can not replace "exact security" or "concrete security".

The exact security is practice-oriented, and aims to give the more precise estimate of running time of an attack task [9], which indicates that one can quantify the security by computing precise bounds on computational effort, rather than an asymptotic bound which is guaranteed to act for a sufficiently large value of the security dominant parameter.

The exact security is obtained through security analysis when a value of a security dominant parameter is given. Security analysis cannot attempt exhaustively all potential attack methods sometimes, but it must consider the most efficient attacks so far [10]. Besides, security analysis does not exclude formal proofs such as the random oracle model and polynomial time Turing reduction.

In general, the security analysis of primitive problems — an interger factorization problem (IFP) and a discrete logarithm problem (DLP) for example is very arduous because it is impossible to search exhaustively all solution methods. However, the security analysis of some composite problems — a multivariate permutation problem (MPP) and an anomalous subset product problem (ASPP) for example is comparatively easy because the combinations of multiple variables in the composite problems may be seached exhaustively on the assumption that the primitives IFP, DLP etc can be solved in tolerable subexponential time [11] [12].

Why are the composite problems MPPand ASPP very hard? There are two reasons: ① finding a specific permutation of the multiple variables in the MPP is very difficult; ② ASSP is proven to be NP-Complete, and verified with experiments to be asymptotically secure.

## 4 Is Cryptology an Academic or Technologic Thing

The termination "-logy" indicates that cryptology inclines towards technology as a cryptosystem may be protected by a patent, and is a type of bit magic to a great extent, which, of course, does not exclude cryptology from bearing scientific ingredients — computational complexity theory and formal security proof in public key cryptography for example. Technology and science never repel each other.

Applied academic things must be checked and verified with practices or experiments as technologic things. The provable security only give a formal verification of security of a public key cryptosystem, but the ultimate and summit verification is practices and experiments, and moreover the experiments on academic things should be repeatable.

Academic things should serve technologic things, but not dissociate from technologic things, further not tempt cryptology to deviate from the technologic criterion, and especially not suppress technologic things.

Therefore, people should appraise a public key encryption scheme or a signature scheme objectively, dialectically, and materialistically, but not subjectively, metaphysically, or idealistically.

## 5  A New Approach to Proof of P ≠ NP

In [13], we prove that the TLP $y \equiv x^x$ (% $M$) with $M$ prime is computationally harder than the DLP $y \equiv g^x$ (% $p$) through asymptotic granularity reduction, which indicates that P ≠ NP holds.

However, the asymptotic granularity reduction is based on the assumption that the noninvertibility of a univariate increasing function $y = f(x)$ with $x > 0$ is in direct proportion to its growth rate reflected by its derivative.

Therefore, the proof of P ≠ NP for which a reward of $1,000,000 is offered by CIM is equivalent to the proof of the above assumption. The latter seems to be easier.

## 6  A Reword Is Offered for Subexponential Time Solutions to the Three REESSE1+ Problems

It may be regarded as a type of proof by experiment.

Here, $n \geq 80$ is the length of a binary string $b_1…b_n \neq 0$ of which the bit-pair string is $B_1…B_{n/2}$ containing at most $n/4$ 00-pairs, the sign % denotes 'modulo', $\overline{M}$ means '$M - 1$' with $M$ prime, and lg $x$ means the logarithm of $x$ to the base 2.

The analysis in [11] and [12] shows that any effectual attack on REESSE1+ will be reduced to the solution of four intractabilities: a multivariate permutation problem (MPP), an anomalous subset product problem (ASPP), a transcendental logarithm problem (TLP), and a polynomial root finding problem (PRFP) so far.

It is well known that it is infeasible in subexponential time to find a large root to the PRFP $ax^n + bx^{n-1} + cx + d \equiv 0$ (% $M$) with $a \notin \{0, 1\}$, $|b| + |c| \neq 0$, $d \neq 0$, and $n, M$ large enough [13][14].

Let $n=80, 96, 112, 128$ with $\lceil \lg M \rceil =384, 464, 544, 640$ for the optimized REESSE1+ encryption scheme or with $\lceil \lg M \rceil =80, 96, 112, 128$ for the lightweight REESSE1+ signing scheme.

Assume that $(\{C_1, …, C_{3n/2}\}, M)$ is a public key, and $(\{A_1, …, A_{3n/2}\}, \{\ell(1), …, \ell(3n/2)\}, W, \delta, M)$ with $W, \delta \in (1, \overline{M})$, $A_i \in \{2, 3, …, 1201\}$, and $\ell(i) \in \{+/-5, +/-7, …, +/-(2(3n/2) + 3)\}$ is a private key, where the sign +/– means that the plus sign + or minus sign – is selected, and unknown to the masses.

The authors promise solemnly that

① anyone who can extract the original private key definitely from the MPP
$$C_i \equiv (A_i W^{\ell(i)})^\delta \text{ (\% } M\text{) for } i = 1, …, 3n/2$$
in DLP subexponential time will be awarded $100000 when $n = 80, 96, 112, 128$ with $\lceil \lg M \rceil = 384, 464, 544, 640$, or $10000 with $\lceil \lg M \rceil = 80, 96, 112, 128$;

② anyone who can recover the original plaintext $b_1…b_n$ definitely from the ASPP
$$\bar{G} \equiv \prod_{i=1}^{n/2}(C_{3(i-1)+B_i})^{\mathcal{B}_i} \text{ (\% } M\text{) with } C_0 = 1 \text{ and } \mathcal{B}_i \text{ a bit-pair shadow}$$
in DLP subexponential time will be awarded $100000 when $n = 80, 96, 112, 128$ with $\lceil \lg M \rceil = 384, 464, 544, 640$, or $10000 with $\lceil \lg M \rceil = 80, 96, 112, 128$, where $\mathcal{B}_i = 0$ if $B_i = 00$, $= 1 +$ the number of successive 00-pairs before $B_i$ if $B_i \neq 00$, or $= 1 +$ the number of successive 00-pairs before $B_i +$ the number of successive 00-pairs after the rightmost non-00-pair if $B_i$ is the leftmost non-00-pair as $b_1…b_{12} = 010000110100 = B_1…B_6 = 01\ 00\ 00\ 11\ 01\ 00$ with $\mathcal{B}_1…\mathcal{B}_6 = 2\ 0\ 0\ 3\ 1\ 0$.

③ anyone who can find the original large answer $x \in (1, \overline{M})$ definitely to the TLP
$$y \equiv (gx)^x \text{ (\% } M\text{)}$$
with known $g, y \in (1, \overline{M})$ in DLP subexponential time will be awarded $100000 when $n = 80, 96, 112,$

128 with $\lceil \lg M \rceil$ = 384, 464, 544, 640, or $10000 with $\lceil \lg M \rceil$ = 80, 96, 112, 128.

Of course, any solution must be described with a formal process, and can be verified with our examples. The time of solving a problem should be relevant to arithmetic steps, but irrelevant to CPU speeds.

The DLP subexponential time means the running time of an algorithm for solving the DLP in the prime field $\mathbb{GF}(M)$ through Index-calculus method at present, namely $L_M[1/3, 1.923]$.

Note that the TLP is written as $y \equiv (gx)^x$ (% $M$) instead of $y \equiv x^x$ (% $M$) due to the asymptotic property of $M$, and in [11] and [12], some pieces of evidence incline people to believe that the subset product problem (SPP) $\bar{G}_1 \equiv \prod_{i=1}^{n} C_i^{b_i}$ (% $M$) is harder than the DLP asymptotically, but due to $\lceil \lg M \rceil \leq 640$ and the density of a related knapsack being low, SPP can almost be solved in DLP subexponential time [12].

## Appendix A ― Computation of Density of a Knapsack from ASPP

### 1) Wrong Computation of Density in Section 5.2 of [12]

It is known from Section 3.2 of [12] that a ciphertext is an ASPP $\bar{G} \equiv \prod_{i=1}^{n} C_i^{b_i}$ (% $M$).

Let $C_1 \equiv g^{u_1}, \ldots, C_n \equiv g^{u_n}, \bar{G} \equiv g^v$ (% $M$), where $g$ is a generator of $(\mathbb{Z}_{M}^{*}, \cdot)$ randomly selected.

Then, seeking $b_1 \ldots b_n$ from $\bar{G}$ is equivalent to solving the congruence

$$u_1 b_1 + \ldots + u_n b_n \equiv v \ (\% \ \overline{M}), \tag{1}$$

where $\{u_1, \ldots, u_n\}$ is called a compact sequence (knapsack) due to $b_i \in [0, n/2+1]$. Seeking $b_1 \ldots b_n$ from (1) is called the anomalous subset sum problem (ASSP).

Note that ① we stipulate that $b_1 \ldots b_n \neq 0$ contains at most $n/2$ 0-bits; ② if $g$ is different, $\{u_1, \ldots, u_n\}$ will be different for the same $\{C_1, \ldots, C_n\}$, and thus $\{u_1, \ldots, u_n\}$ has randomicity.

When (1) will be reduced through the LLL lattice basis reduction algorithm, it should be converted into a non-modular form:

$$b_1 u_1 + \ldots + b_n u_n \equiv v + k\overline{M}, \tag{2}$$

where $k \in [0, n]$ is an integer. To seek the original solution to (2), $k$ must traverse from 0 to $n$.

Let $D$ be the density of the compact sequence $\{u_1, \ldots, u_n\}$. We see that in Section 5.2 of [12], the formula $D \approx n^2 / \lceil \lg M \rceil$ is wrong, which is first pointed out by Xiangdong Fei (an associate professor from Nanjing University of Technology).

### 2) Right Computation of Density in Section 5.2 of [12]

Considering the structure of a lattice basis from (1) and the bit-length of a bit shadow $b_i \in [0, n/2+1]$ (on the assumption that $b_1 \ldots b_n$ contains at most $n/2$ 0-bits), the right computation of density of an ASSP knapsack in [12] should be

$$D = \sum_{i=1}^{n} \lceil \lg(n/2+1) \rceil / \lceil \lg M \rceil = n \lceil \lg(n/2+1) \rceil / \lceil \lg M \rceil.$$

Concretely speaking,

for $n$ = 80 with $\lceil \lg M \rceil$ = 696, $D$ = 80 × 6 / 696 ≈ 0.6897 < 1;
for $n$ = 96 with $\lceil \lg M \rceil$ = 864, $D$ = 96 × 6 / 864 ≈ 0.6667 < 1;
for $n$ = 112 with $\lceil \lg M \rceil$ = 1030, $D$ = 112 × 6 / 1030 ≈ 0.6524 < 1;
for $n$ = 128 with $\lceil \lg M \rceil$ = 1216, $D$ = 128 × 7 / 1216 ≈ 0.7368 < 1.

These densities mean that the original solution to (1) may possibly be found through LLL lattice basis reduction (not certainly, and even with very low probability) because $D < 1$ only assure that the shortest vector is unique, but it cannot assure that the vector of the original solution is just the shortest vector in the reduced basis.

### 3) Bit Shadows Enhance Resistance of a Low Density ASSP Knapsack to Attacks

The LLL algorithm is to reduce a lattice basis $\langle 1, 0, \ldots, 0, \tilde{N}u_1 \rangle$, $\langle 0, 1, \ldots, 0, \tilde{N}u_2 \rangle$, ..., $\langle 0, 0, \ldots, 1, \tilde{N}u_n \rangle$, $\langle 1/2, 1/2, \ldots, 1/2, \tilde{N}(v + k\overline{M}) \rangle$, where $\tilde{N} > 1/2(n)^{1/2}$. No matter whether $(v + k\overline{M})$ is a classical subset sum or an anomalous subset sum, and whether the density is less than 1 or greater than 1, the LLL algorithm runs by its inherent rules. Lastly, the $n + 1$ vectors $\langle \hat{e}_1, \ldots, \hat{e}_n, \hat{e}_{n+1} \rangle$'s which occur in the reduced basis are the first $n + 1$ approximately shortest vectors, including the shortest vector, of which quite some satisfy $\hat{e}_1 u_1 + \ldots + \hat{e}_n u_n \equiv v + k\overline{M}$ (omitting the term $\hat{e}_{n+1} = 0$). If $D < 1$, the shortest vector is unique; and if $(v + k\overline{M})$ is a classical subset sum, the shortest vector is just the original solution.

We know from the above discussion that it has two necessary conditions to solve a SSP or ASSP

through LLL lattice basis reduction: ① the vector of the original solution is the shortest; ② the shortest vector in the lattice is unique. $D < 1$ assures that the shortest vector is unique; and a classical subset sum assures that the vector of the original solution is just the shortest vector.

Return to (1). Even though the density of an ASSP knapsack is less than 1, the **original** solution is not necessarily found since $b_i \in [0, n/2+1]$ is a bit shadow (indicates that the **original** solution does not necessarily occur in the reduced basis), and there likely exist many solutions in the lattice.

For example, let $n = 4$ (short but without loss of generality), $M = 263$, $\{u_1, \ldots, u_4\} = \{48, 71, 257, 4\}$, and $v = 261$ ($u_i$ and $v$ are obtained through the discrete logarithms of $C_i$ and $\bar{G}$). Here, the density of $\{u_1, \ldots, u_4\}$ is $D = n \lceil \lg(n/2+1) \rceil / \lceil \lg M \rceil = 4 \times 2 / 9 = 0.8889 < 1$.

Assume that a plaintext $b_1 \ldots b_4 = 1100$, and its related bit shadow string $\bar{b}_1 \ldots \bar{b}_4 = 1300$ (thus $1 \times 48 + 3 \times 71 = 261$).

However, according to LLL lattice basis reduction, sought solution will be 0011 (notice, not the original) because there is $1 \times 257 + 1 \times 4 = 261$, and $\langle 0, 0, 1, 1 \rangle \mid (1^2 + 1^2)^{1/2}$ is the shortest vector in the lattice while $\langle 1, 3, 0, 0 \rangle \mid (1^2 + 3^2)^{1/2}$ is not the shortest, and even it will not occur in the reduced basis consisting of 5 vectors.

### 4) Density of an Optimized ASSP Knapsack

The modulus of prototypal REESSE1+ is relatively large, so in practice, it needs to be optimized (optimized REESSE1+ is called JUNA).

Return to Section 6. When $n = 80, 96, 112, 128$, correspondingly there is $\lceil \lg M \rceil = 384, 464, 544, 640$. The density of an optimized ASSP knapsack is $D = (3n/2) \lceil \lg(n/4+1) \rceil / \lceil \lg M \rceil$ with $B_i \in [0, n/4+1]$ (on the assumption that $B_1 \ldots B_{n/2}$ contains at most $n/4$ 00-pairs).

Concretely speaking,
for $n = 80$ with $\lceil \lg M \rceil = 384$, $D = 120 \times 5 / 384 \approx 1.5625 > 1$;
for $n = 96$ with $\lceil \lg M \rceil = 464$, $D = 144 \times 5 / 464 \approx 1.5517 > 1$;
for $n = 112$ with $\lceil \lg M \rceil = 544$, $D = 168 \times 5 / 544 \approx 1.5441 > 1$;
for $n = 128$ with $\lceil \lg M \rceil = 640$, $D = 192 \times 6 / 640 \approx 1.8000 > 1$.

Under the circumstances, owing to $D > 1$ (indicates there will exist many solutions to the ASSP, and even the shortest vector is also nonunique), it is impossible to find the original plaintext $b_1 \ldots b_n$ through LLL lattice basis reduction. Therefore, at present there exists no subexponential time solution to the ASPP used in the optimized encryption scheme.